\documentclass[twocolumn]{aastex63}
\usepackage{natbib}
\usepackage{amsmath}
\usepackage{chngcntr}
\usepackage{hyperref}
\usepackage[figuresright]{rotating}
\usepackage{ctable}
\usepackage{float}
\usepackage{graphicx}
\usepackage{realboxes}
\usepackage{epsfig}
\usepackage{threeparttablex, tablefootnote}
\usepackage{subfigure}
\usepackage{graphicx}
\newcommand{\uat}[2]{\href{http://astrothesaurus.org/uat/#2}{#1 (#2)}}
\usepackage{lineno}
\usepackage{CJK}

\newcommand{\cahk}{Ca \scriptsize{\uppercase\expandafter{\romannumeral2}} \normalsize H$\&$K }

\shortauthors{Han et al.}

\begin{document}

\begin{CJK*}{UTF8}{gbsn}
\title{Revisiting the activity--rotation relation for evolved stars}


\author{Henggeng Han}
\affil{Key Laboratory of Optical Astronomy, National Astronomical Observatories, Chinese Academy of Sciences, Beijing 100101, People's Republic of China}

\author{Song Wang$^\dagger$}
\affil{Key Laboratory of Optical Astronomy, National Astronomical Observatories, Chinese Academy of Sciences, Beijing 100101, People's Republic of China}
\affil{Institute for Frontiers in Astronomy and Astrophysics, Beijing Normal University, Beijing, 102206, People's Republic of China}
\email{$^\dagger$ Corresponding Author: songw@bao.ac.cn}

\author{Xue Li}
\affil{Key Laboratory of Optical Astronomy, National Astronomical Observatories, Chinese Academy of Sciences, Beijing 100101, People's Republic of China}
\affil{School of Astronomy and Space Science, University of Chinese Academy of Sciences, Beijing 100049, People's Republic of China}

\author{Chuanjie Zheng}
\affil{Key Laboratory of Optical Astronomy, National Astronomical Observatories, Chinese Academy of Sciences, Beijing 100101, People's Republic of China}
\affil{School of Astronomy and Space Science, University of Chinese Academy of Sciences, Beijing 100049, People's Republic of China}

\author{Jifeng Liu}
\affil{Key Laboratory of Optical Astronomy, National Astronomical Observatories, Chinese Academy of Sciences, Beijing 100101, People's Republic of China}
\affil{School of Astronomy and Space Science, University of Chinese Academy of Sciences, Beijing 100049, People's Republic of China}
\affil{Institute for Frontiers in Astronomy and Astrophysics, Beijing Normal University, Beijing, 102206, People's Republic of China}
\affil{New Cornerstone Science Laboratory, National Astronomical Observatories, Chinese Academy of Sciences, Beijing, 100012, People's Republic of China}

\begin{abstract}

The magnetic dynamo mechanism of giant stars remains an open question, which can be explored by investigating their activity--rotation relations with multiple proxies.
By using the data from the LAMOST and \emph{GALEX} surveys, we carried out a comprehensive study of activity--rotation relations of evolved stars based on \cahk lines, $\rm{H\alpha}$ lines and near ultraviolet (NUV) emissions. 
Our results show that evolved stars and dwarfs obey a similar power-law in the unsaturated region of the activity--rotation relation, indicating a common dynamo mechanism in both giant and dwarfs.
There is no clear difference in the activity levels between red giant branch stars and red clump stars, nor between single giants and those in binaries. 
Additionally, our results show that the NUV activity levels of giants are comparable to those of G- and K-type dwarfs and are higher than those of M dwarfs.

\end{abstract}

\section{Introduction} 
Stellar magnetic fields are generated and maintained by stellar rotation and turbulent motion, named magnetic dynamo \citep{1955ApJ...122..293P}. Stellar chromospheric activities are mostly caused by the variation of their magnetic fields, thus chromospheric emissions are excellent tracers of stellar magnetic fields \citep{1978ApJ...226..379W, 1985ApJ...295..162B}. The widely used chromospheric activity indices including \cahk lines, $\rm{H\alpha}$ lines and ultraviolet (UV) fluxes, could map the activities of different layers of the chromosphere \citep{1981ApJS...45..635V, 2015ApJ...809..157F, 2017ARA&A..55..159L}.

Magnetic activities were observed among late-type dwarfs across the Hertzsprung--Russell(HR) diagram. Pioneer works have suggested that there is a dividing-line in the HR diagram, which separates two distinct groups of stars, i.e., a solar-type group with detected magnetic activities and a non-solar-type group corresponding to cool giants, among which magnetic activities were not observed \citep{1979ApJ...229L..27L, 1981ApJ...250..293A}. However, later on, many studies have found signals of magnetic activities from subgiants and giants \citep[e.g.][]{1983ApJ...271..672B, 1989ApJ...346..303S}. 
Investigating the magnetic activity of evolved stars has gradually became an important topic. 

As a star evolves it will gradually lose angular momentum and thus present weak magnetic activity \citep{1972ApJ...171..565S}. However, recent researches have found many active evolved stars with small rotation periods \citep{2020A&A...639A..63G, 2022A&A...668A.116G}.
Some of them are in close-in binaries and have been spun-up by tidal interaction of their companions \citep{2015ApJ...807...82T}, while there is still a large fraction of them are single stars. The origin of magnetic activity presented on these stars is worth investigating.

One of the open questions regarding on stellar activity of giants is, could main sequence and evolved stars develop a similar stellar dynamo? Similar to dwarfs, the magnetic activities of giants have been found to be positively correlated to rotation velocity \citep[e.g.][]{1994A&A...281..855S}. Recent works have found that main-sequence and evolved stars fall on the same rotation–activity sequence in the unsaturated regime \citep{2020NatAs...4..658L, 2020ApJ...902..114W}. However, their samples mainly contain subgiants.  
\cite{2020AJ....160...12D} proposed that for evolved stars, the NUV activity-rotation relation of evolved stars is similar to that of M dwarfs but have different slope in the decay region. Whether the evolved stars and dwarfs obey the same activity--rotation relation is still unclear, and a large sample of evolved star with multi-band observations is needed to revisit this question.

Nowadays there are many elite sky surveys including the \emph{Kepler} mission \citep{2010ApJ...713L..79K}, the {\it Galaxy Evolution Explorer} (\emph{GALEX}) mission \citep{2007ApJS..173..682M}, and the Large Sky Area Multi-Object Fiber Spectroscopic Telescope (LAMOST)-\emph{Kepler}/K2 projects \citep{2020RAA....20..167F}, making it possible to carry out a comprehensive study of chromospheric activity of evolved stars. 
In this work, based on the catalog from \cite{2017A&A...605A.111C}, which measured rotation periods for giant stars, we investigated stellar activity--rotation relations of evolved stars, including red giant branch (RGB) and red clump (RC) stars. This paper is organized as follows. Section 2 describes the sample construction and methods. Section 3 gives the results and discussions.

\section{Sample construction and methods}
\subsection{Sample construction}

\begin{figure*}[t]
\includegraphics[width=0.95\textwidth]{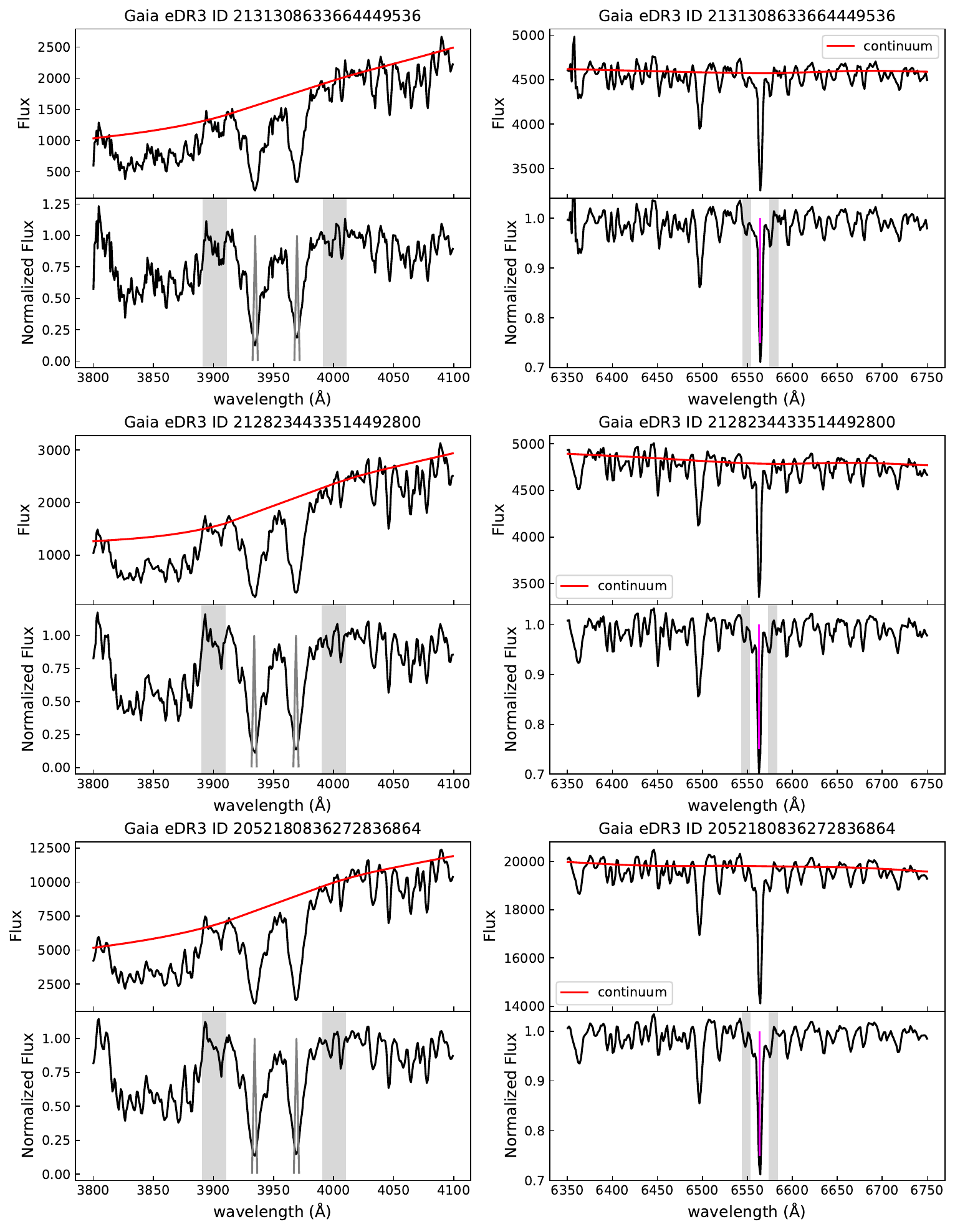}
\caption{Left panels: Examples of LAMOST blue-band spectra and their normalization. Red lines are continua used for normalization. Grey shaded areas mark the continua used for the calculation of $S_{\rm{LAMOST}}$ using the \cahk lines. Grey triangles are integration windows of \cahk lines with an FWHM of 2.18 \AA. Right panels: Examples of LAMOST red-band spectra and their normalization. Grey shaded areas mark the continua used for the calculation of $EW_{\rm{H\alpha}}$ using ${\rm H\alpha}$ lines. Magenta lines mark the RV-corrected line centers of $\rm{H\alpha}$ lines.}
\label{spectrum}
\end{figure*}

The \emph{Kepler} mission \citep{2010ApJ...713L..79K} has provided us with roughly 200,000 light curves with high quality, based on which \cite{2017A&A...605A.111C} have measured rotation periods of 361 giants, including RGB and RC stars. Various contamination including pulsation and binarity have been excluded, making this sample an ideal one to study the relation between activity and rotation of 
evolved stars. We cross-matched the sample with LAMOST DR10 to obtain their spectra and stellar parameters \citep{2012RAA....12.1197C, 2015RAA....15.1095L}. For targets with multiple observations, the spectrum with the highest signal-to-noise ratio and corresponding parameter estimation were used. 

The revised \emph{GALEX} catalog GR6+7 contains 292,296,11 targets with near ultraviolet (NUV) and far ultraviolet (FUV) observations \citep{2017ApJS..230...24B}. 
We also cross-matched the sample from \cite{2017A&A...605A.111C}, which have LAMOST stellar parameters, with the \emph{GALEX} catalog using a radius of 3''. In order to avoid contamination, we removed targets with ``nuv$\_$artifact'' larger than one, which represents ``NUV bright star window reflection'', ``dichroic reflection'', ``detector run proximity'' or ``bright star ghost''. This led to 86 targets with UV observations. All the \emph{GALEX} data could be found at MAST \citep{https://doi.org/10.17909/t9h59d}. 

Generally, giants are thought to have low magnetic activity levels. However, light curves from \emph{Kepler} observations show that roughly 8$\%$ red giants exhibit rotational modulation, suggesting notable magnetic activity \citep[e.g.][]{2020A&A...639A..63G}. Meanwhile, the authors suggested that about 15$\%$ of these red giants belong to close-in binaries (according to their rotation periods), which may exhibit higher magnetic activity compared to single red giants. This enhancement may bias the activity--rotation relation for single stars.

Therefore, we took additional steps to identify potential binaries. First, we cross-matched the giant sample with LAMOST DR10 to obtain radial velocity (RV) values \citep{2015RAA....15.1095L}, and marked giants with RV variations larger than 10 km/s. Second, we cross-matched the giants with $Gaia$ eDR3 and the None-Single Star (NSS) catalog \citep{2021A&A...649A...1G}, marking those with RUWE values larger than 1.4 or listed in the NSS catalog. Third, we checked the giants against the SIMBAD database, but none were marked as binaries. Therefore, we classified our sample into single stars and binaries.

Moreover, we cross-matched our sample with Gaia eDR3 catalog \citep{2021A&A...649A...1G} to get Gaia \emph{G}, \emph{BP} and \emph{RP} photometry. We adopted the Gaia eDR3 distances from \cite{2021AJ....161..147B}. \emph{B} band and \emph{V} band photometry were extracted from the UCAC4 catalog \citep{2013AJ....145...44Z}. Finally, the interstellar reddening were derived based on the Pan-STARRS1 3D dust map \citep{2018JOSS....3..695M, 2019ApJ...887...93G}. 
\begin{figure}
\centering
\includegraphics[width=0.45\textwidth]{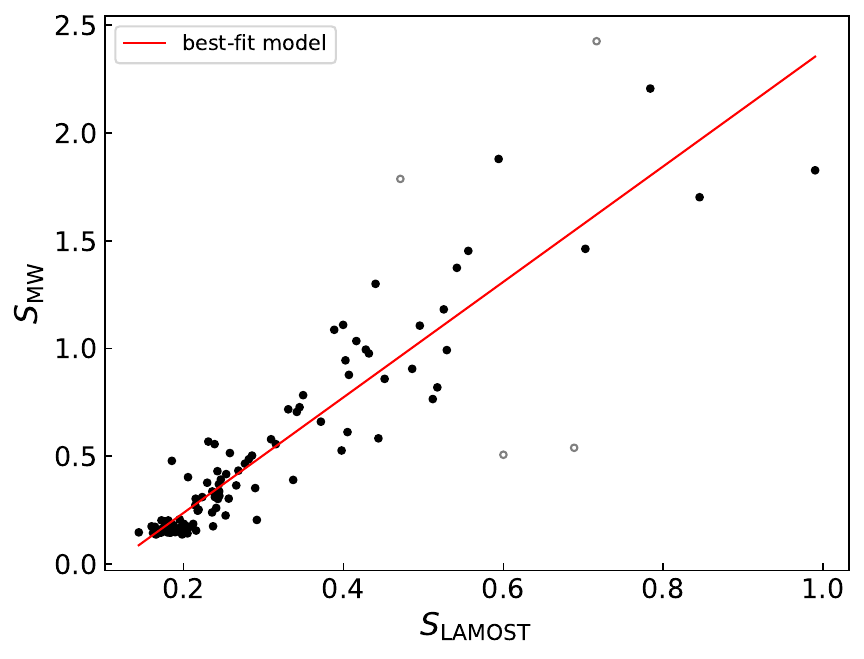}
\caption{Relation between $S_{\rm{LAMOST}}$ and $S_{\rm{MW}}$. The red line represents the best-fit model. The gray open circles are not included in the fitting.}
\label{relation_S}
\end{figure}

\subsection{LAMOST spectra}

LAMOST is a reflecting Schmidt telescope, with an aperture of 3.6 to 4.9 m and a field of view of 5 degrees \citep{2012RAA....12.1197C}. Two kinds of spectra with different resolution have been provided. One is the low-resolution spectra with a resolution of R$\sim$1800, the other is the medium-resolution spectra with a resolution of R$\sim$7500. The low-resolution spectra cover a wavelength range of 3700--9000 \AA, while the median-resolution spectra have a wavelength coverage of 4950--5350 \AA\ at the blue arm and 6300--6800 \AA\ at the red arm \citep{2020RAA....20..167F}. Since we focused on both the \cahk and $\rm{H\alpha}$ lines, we used the low-resolution spectra in this work.

Before the calculation of activity indices, we corrected the RVs of the spectra using the RV values provided by the LAMOST DR10 catalog \citep{2015RAA....15.1095L}. Meanwhile, we used the smoothing spline method from the \emph{laspec} package \citep{2020ApJS..246....9Z} to normalize the LAMOST spectra. 
Figure \ref{spectrum} shows some examples of normalized spectra, the \cahk and $\rm{H\alpha}$ lines and corresponding continua. 
To guarantee the quality of LAMOST spectra, in following analysis we excluded those with a signal-to-noise ratio lower than 30.


\subsection{Rotation periods and Rossby number}
As discussed above, we used the rotation periods estimated from $Kepler$ data \citep{2017A&A...605A.111C}. Considering that the \emph{Kepler} satellite would rotate 90 degrees every 90 days \citep{2016ksci.rept....1V}, which could cause systematic variability and contaminate light curves, we removed all the targets with rotation periods between 85 days and 95 days. 

To calculate the Rossby number (Ro = $P_{\rm{rot}} / \tau_{\rm{c}}$), we further calculated the convective turnover time ($\tau_{\rm{c}}$) using the Yale-Potsdam Stellar Isochrones \citep[YAPSI;][]{2017ApJ...838..161S}. The subgrids that contain solar calibration were adpoted.
First, we calculated the bolometric luminosity of our sample, $L_{\rm{bol}} = 10^{-0.4\times(m_{\lambda}-5\rm{log}_{10}\it{d}+5-\it{A}_{\lambda}+\it{BC}_{\lambda}-\it{M}_{\odot})}L_{\odot}$, in \emph{G, BP, RP, J, H} and \emph{Ks} bands. Here $M_{\rm{\odot}} = 4.74$ mag is the solar bolometric magnitude and $L_{\rm{\odot}} = 3.828 \times 10^{33}$ erg/s is the solar bolometric luminosity. We used the mean and standard deviation of $L_{\rm{bol}}$ corresponding to six bands as the final $L_{\rm{bol}}$ and its error. Second, for each target, the metallicity gird closest to the observed [Fe/H] was chosen and the models within the typical errors of $T_{\rm{eff}}$ and log$L_{\rm{bol}}$ were picked out, which were set to be 200 K and 0.2, respectively. We then calculated the median value of $\tau_{c}$ with these models as the final $\tau_{c}$.

\subsection{Activity indices}
\subsubsection{\rm{\cahk} $index$: $R_{\rm{HK}}^{'}$}
\label{cahk.sec}

Known as the Wilson-Bappu effect, the widths of the emission cores of \cahk lines would increase towards higher luminosities, especially for evolved stars \citep{1957ApJ...125..661W}. Consequently, the Mount Wilson Observatory (MWO) used a 2 \AA \, slit to estimate the fluxes of the \cahk lines of giant stars \citep{1991ApJS...76..383D}. Recently, this effect has been revisited by \cite{2018MNRAS.480.2137S} to illustrate that when calculating $S$-index using the data from spectra, a wider integration window is also needed for giants.

To quantify the strength of \cahk lines, we first calculated the \emph{S}-index using the normalized LAMOST spectra:
\begin{equation}
    S_{\rm{LAMOST}} = 8 \cdot \alpha \cdot \frac{H + K}{R + V}.
\end{equation}
Unlike previous studies which used an integration window with a full width at half maximum (FWHM) of 1.09 \AA \, \citep[e.g.][]{2016NatCo...711058K, 2020ApJS..247....9Z}, in this work, we used a triangle bandpass with an FWHM of 2.18 \AA, consistent with observations of the MWO. \emph{H} and \emph{K} are the integrated flux in the triangle bandpass centered at 3969.59 \AA \, and 3934.78 \AA, respectively, as shown in Figure \ref{spectrum} (left panels). \emph{R} and \emph{V} represent the integrated flux in the 20 \AA \, rectangle windows (shaded areas in panel (a) of Figure \ref{spectrum}) centered at 4001 \AA \, and 3901 \AA, respectively. The $\alpha$ was set to be 1.8 according to \cite{2007AJ....133..862H}. The factor 8 accounts for the different exposing strategies of different channels of Mount Wilson spectrophotometer. In order to derive the error of $S_{\rm{LAMOST}}$, for each target, 1000 synthetic spectra were simulated based on the Gaussian noise at each wavelength. We then calculated $S_{\rm{LAMOST}}$ of each spectrum and used their standard deviation as the error of $S_{\rm{LAMOST}}$. 

To convert $S_{\rm{LAMOST}}$ to the well-known Mount Wilson S-index ($S_{\rm{MW}}$), we cross-matched the sample from \cite{2018A&A...616A.108B} with LAMOST DR10 and a linear model was fitted to the common targets using the least squares:
\begin{equation} 
    \emph{S}_{\rm{MW}} = a \times \emph{S}_{\rm{LAMOST}} + b,
\end{equation}
where $a =$ 2.68$\pm$0.23 and $b = -$0.3$\pm$0.06 (Figure \ref{relation_S}). To estimate the uncertainty of the linear fitting, we repeated the fitting process 10,000 times through random sampling the errors in $S_{\rm{LAMOST}}$. The standard deviations of the 10,000 values of $a$ and $b$ were then used as the uncertainties of $a$ and $b$, respectively. These uncertainties were subsequently incorporated into the errors of $R_{\rm{HK}}^{'}$ through error propagation.

\begin{figure}
\subfigure[]{
\includegraphics[width=0.45\textwidth]{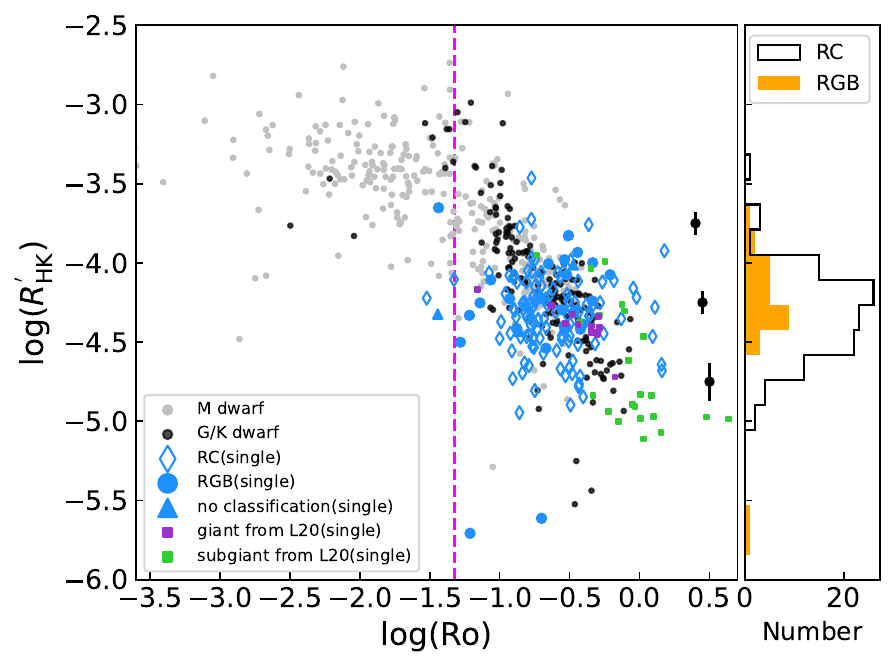}}
\subfigure[]{
\includegraphics[width=0.45\textwidth]{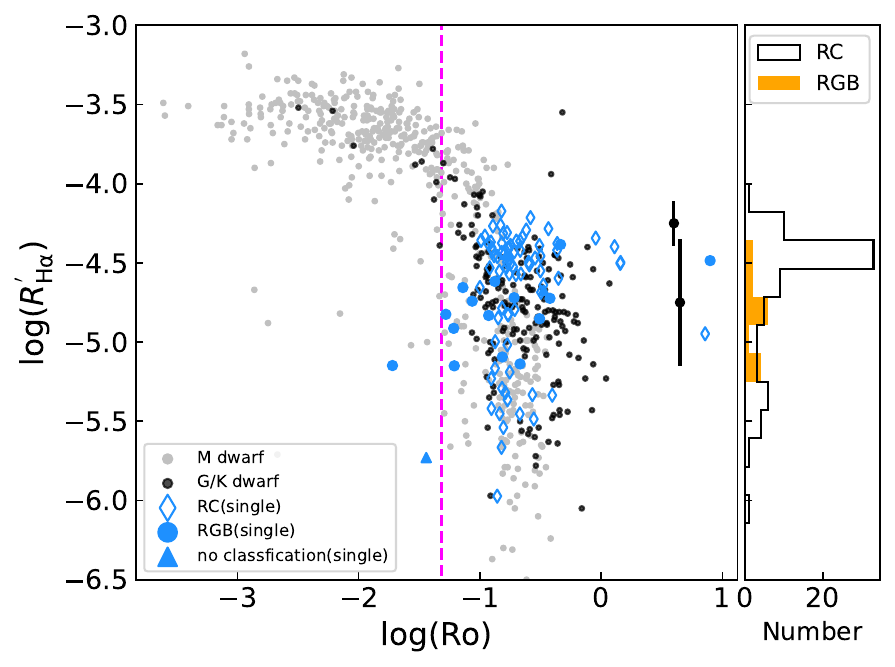}}
\centering
\subfigure[]{
\includegraphics[width=0.45\textwidth]{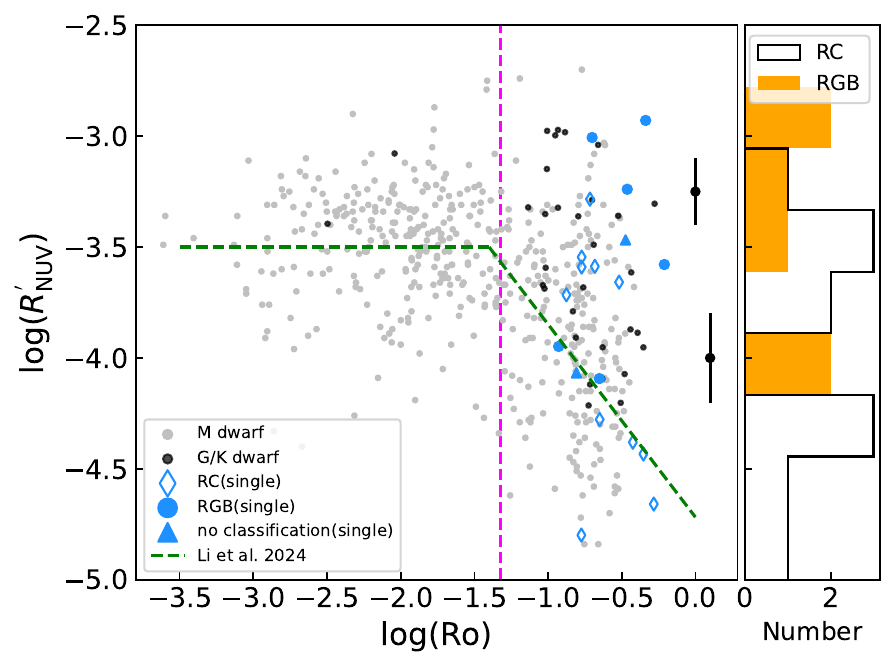}}
\caption{Activity--rotation relations for single stars including $R_{\rm{HK}}^{'}-$Ro relation (Panel (a)), $R_{\rm{H\alpha}}^{'}-$Ro relation (Panel (b)), and $R_{\rm{NUV}}^{'}-$Ro relation (Panel (c)). 
The black segments represent typical errors of activity indices for giants at different activity levels. The magenta lines mark the knee point, log(Ro) $=$ $-$1.32, separating the saturated and the unsaturated regimes from \cite{2024ApJS..273....8H}. The green dashed line is the fitting result for M dwarfs from \cite{2024ApJ...966...69L}. }
\label{relation}
\end{figure}

\begin{figure}
\subfigure[]{
\includegraphics[width=0.45\textwidth]{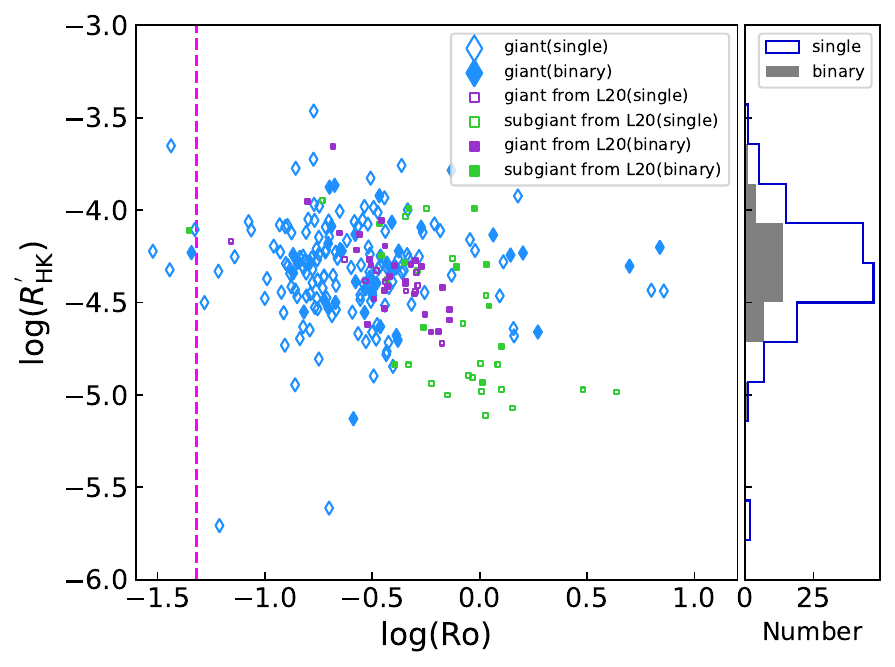}}
\subfigure[]{
\includegraphics[width=0.45\textwidth]{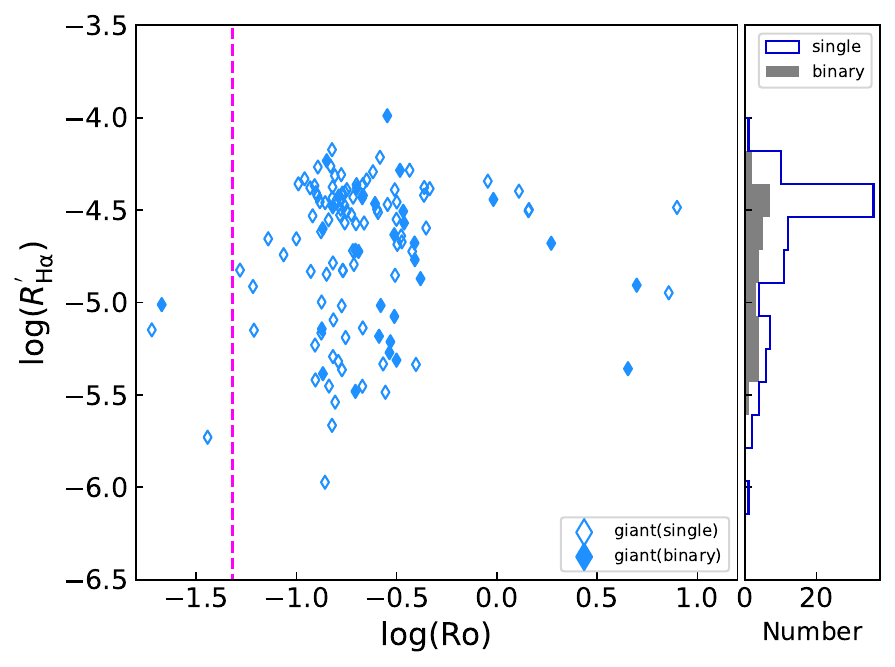}}
\centering
\subfigure[]{
\includegraphics[width=0.45\textwidth]{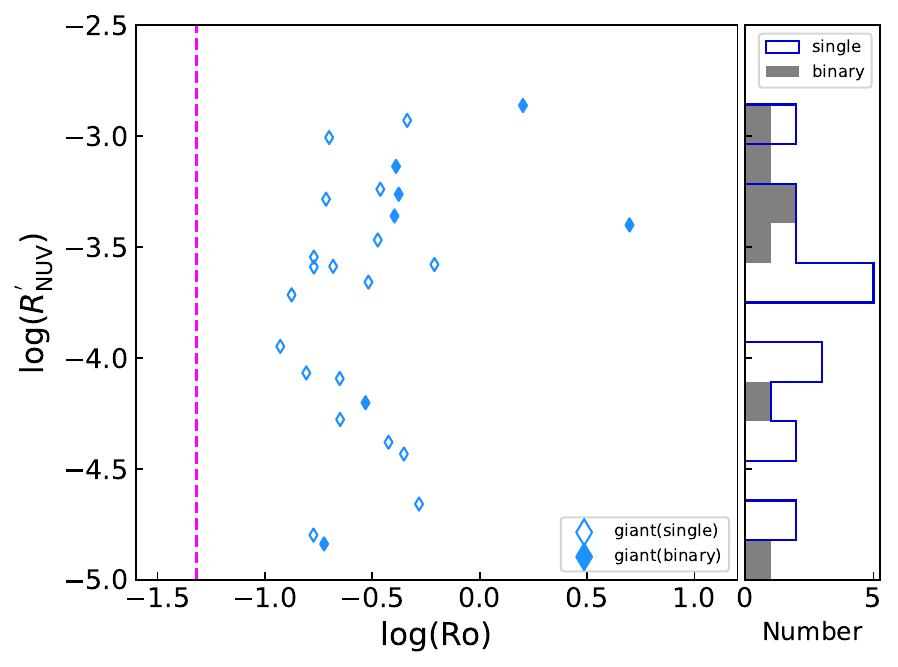}}
\caption{Activity--rotation relations of single and binary evolve stars, including $R_{\rm{HK}}^{'}-$Ro relation (Panel (a)), $R_{\rm{H\alpha}}^{'}-$Ro relation (Panel (b)), and $R_{\rm{NUV}}^{'}-$Ro relation (Panel (c)). The magenta lines mark the knee point, log(Ro) $=$ $-$1.32, separating the saturated and the unsaturated regimes from \cite{2024ApJS..273....8H}.}
\label{relation_binary}
\end{figure}

It is important to note that the common sources between LAMOST DR10 and \cite{2018A&A...616A.108B} are all dwarfs, and their $S_{\rm{LAMOST}}$ values were estimated using a triangle bandpass with a typical FWHM of 1.09 \AA. We assumed that the conversion derived from dwarfs is also applicable for giants, whose $S_{\rm{LAMOST}}$ values were estimated using a triangle bandpass with an FWHM of 2.18 \AA, this assumption may introduce some uncertainties.

Finally, in order to get the pure chromospheric activity, the contribution of photosphere and dependence on spectral type of \emph{S}-index need to be corrected \citep{1984ApJ...279..763N}. We then followed the approach of \cite{1979ApJS...41...47L}:
\begin{equation}
    R_{\rm{HK}}^{'} = \frac{\mathcal{F}_{\rm{HK}} - \mathcal{F}_{\rm{HK, ph}}}{\sigma T_{\rm{eff}}^{4}},
\end{equation}
where $\mathcal{F}_{\rm{HK}}$ is total flux of \cahk lines, which was calculated as $\mathcal{F}_{\rm{HK}} = S_{\rm{MW}} \cdot \mathcal{F}_{\rm{RV}} / 8\alpha$ according to \cite{1982A&A...113....1M}. Following \cite{2013A&A...549A.117M}, here $\alpha = 2.4$. $\mathcal{F}_{\rm{RV}}$ is the total flux in the \emph{R} and \emph{V} windows. For dwarfs with $B-V$ between 0.44 and 1.6, $\mathcal{F}_{\rm{RV}} = 19.2 \times 10^{8.25-1.67(B-V)}$. For giants with $B-V$ between 0.76 and 1.18, $\mathcal{F}_{\rm{RV}} = 19.2 \times 10^{8.33-1.79(B-V)}$. $\mathcal{F}_{\rm{HK, ph}}$ is the contribution of photosphere to the \cahk lines. For dwarfs with $B-V$ between 0.44 and 1.28, 1.28 and 1.6, $\mathcal{F}_{\rm{HK, ph}} = 10^{7.49-2.06(B-V)}$ and $\mathcal{F}_{\rm{HK, ph}} = 10^{6.19-1.04(B-V)}$, respectively. While for giants with $B-V$ between 0.76 and 1.18, $\mathcal{F}_{\rm{HK, ph}} = 10^{7.61-2.37(B-V)}$. Here $B-V$ is the intrinsic color. $\sigma$ is the Stefan-Boltzmann constant and $T_{\rm{eff}}$ is the effective temperature. The errors of $S_{\rm{MW}}$ were derived through the error propagation of $S_{\rm{LAMOST}}$ errors. Then errors of $R_{\rm{HK}}^{'}$ were estimated through random sampling of errors of \emph{$S_{\rm{MW}}$} and $T_{\rm{eff}}$.

\subsubsection{$\rm{H\alpha}$ $index$: $R_{\rm{H\alpha}}^{'}$}
Following \cite{2023ApJS..264...12H}, we calculated the $\rm{H\alpha}$ activity index $R_{\rm{H\alpha}}^{'}$. First, the equivalent widths of $\rm{H\alpha}$ lines were defined as
\begin{equation}
    EW = \int \frac{F_{\rm{c}} - F_{\rm{\lambda}}}{F_{\rm{c}}}d\lambda.
    \label{ew.eq}
\end{equation}
The integration interval was set to be 20 \AA \, centered at 6564.6 \AA. \, The $F_{\rm{c}}$ was defined as the median value of pseudo-continua (shaded areas in panel (b) of Figure \ref{spectrum}). Similar to the \rm{\cahk} lines, the photospheric contribution to the $\rm{H\alpha}$ lines needs to be subtracted:
\begin{equation}
    EW^{'} = EW - EW_{\rm{ph}},
\end{equation}
where $EW_{\rm{ph}}$ was estimated using Eq. (\ref{ew.eq}) based on PHOENIX synthetic spectra \citep{2013A&A...553A...6H} with the closest $T_{\rm{eff}}$, surface gravity (log$g$) and metallicity ([Fe/H]) to our giant sample. 

Then, following \cite{2004PASP..116.1105W} we converted $EW^{'}$ to normalized $\rm{H\alpha}$ luminosity: 
\begin{equation}
    R_{\rm{H\alpha}}^{'} = L_{\rm{H\alpha}}/L_{\rm{bol}} = \chi \cdot EW^{'}.
\end{equation}
The factor $\chi$ was calculated as
\begin{equation}
    \chi = \frac{f_{\rm{\lambda6564}}}{\sigma T_{\rm{eff}}^{4}}.
\end{equation}
$f_{\rm{\lambda6564}}$ is the continuum flux at 6564 \AA, \, which was estimated through fitting the continuum of PHOENIX spectra \citep{2013A&A...553A...6H} with nearest stellar parameters to our giant sample. The error of $EW$ was calculated with similar steps to \emph{S}$_{\rm LAMOST}$. The error of $R_{\rm{H\alpha}}^{'}$ was estimated using error propagation of $EW$.

\subsubsection{\rm{NUV} $index$: $R_{\rm{NUV}}^{'}$}
\label{nuv.sec}
To quantify NUV activity level, we calculated the $R_{\rm{NUV}}^{'}$ index following \cite{2024ApJ...966...69L}:
\begin{equation}
    R_{\rm{NUV}}^{'} = \frac{f_{\rm{NUV, obs}} \cdot (\frac{d}{R})^{2} - f_{\rm{NUV, ph}}}{\sigma T_{\rm{eff}}^{4}}. 
\end{equation}
Here $f_{\rm{NUV, obs}}$ is the observed NUV flux with interstellar reddening being corrected:
\begin{equation}
\begin{split}
f_{\rm{NUV, obs}} = 10^{-0.4 \cdot (m_{\rm{NUV}} - 20.08 - R_{\rm{NUV}} \cdot E(B-V))} \\
\times \delta \lambda_{\rm{NUV}}\times 2.06\times10^{-16}.
\end{split}
\label{nuvobs.eq}
\end{equation}
$f_{\rm{NUV,ph}}$ is the photospheric contribution to the NUV flux. $R_{\rm{NUV}}$ = 8.71 is the NUV extinction coefficients \citep{1989ApJ...345..245C}. In this work, we applied the PARSEC evolutionary tracks \citep{2012MNRAS.427..127B} with metallicity grids of $-$2.5, $-$2.0, $-$1.5, $-$1.0, $-$0.5, 0.0, 0.3, 0.5, 1.0 to estimate the photospheric contribution in NUV band. For each target, we first selected two model grids with the closest metallicity. Then, we extracted the best model by comparing the theoretical and observed $T_{\rm{eff}}$ and $L_{\rm{bol}}$ to get the photospheric NUV magnitude, which was further converted to $f_{\rm{NUV,ph}}$. Finally, we obtained the final $f_{\rm{NUV,ph}}$ by linear interpolation using metallicity. In Eq. (\ref{nuvobs.eq}), $\delta\lambda_{\rm{NUV}}$ is the effective bandwidth of 1060 \AA \, and $R_{\rm{NUV}}$ is the extinction coefficient in NUV band from \cite{1999PASP..111...63F}. $d$ is the Gaia eDR3 distance \citep{2021AJ....161..147B}. Stellar radius $R$ was calculated as $R = \sqrt{\frac{L_{\rm{bol}}}{4\pi \sigma T_{\rm{eff}}^{4}}} =  \sqrt{\frac{10^{-0.4\times(m_{\lambda}-5\rm{log}_{10}\it{d}+5-\it{A}_{\lambda}+\it{BC}_{\lambda}-\it{M}_{\odot})}L_{\odot}}{4\pi \sigma T_{\rm{eff}}^{4}}}$. $m_{\rm{\lambda}}$, $A_{\lambda}$, $BC_{\rm{\lambda}}$ are the apparent magnitude, extinction, defined as $R_{\rm{\lambda}}\cdot E(B-V)$, and bolometric correction in different bands. We used the $isochrones$ package \citep{2015ascl.soft03010M} to derive the $BC_{\rm{\lambda}}$ based on stellar parameters including $T_{\rm{eff}}$, log$g$ and [Fe/H]. The calculation was repeated for \emph{G, BP, RP, J, H} and \emph{Ks} bands. We used the mean value and standard deviation of $R$ in six bands as the final stellar radius and its error, respectively. The error of $R_{\rm{NUV}}^{'}$ was estimated using the sampling of errors of $f_{\rm{NUV}}$, distance, radius and $T_{\rm{eff}}$.

\section{Results and Discussions}

Stellar activity--rotation relations of dwarfs have been extensively studied. To construct activity--rotation relations of evolved stars and compare them to those of dwarfs, we first gathered samples from \cite{2024ApJ...966...69L} and \cite{2023ApJS..264...12H}, which mainly contain M-, K- and G-type dwarfs with measured Ro and chromospheric activity indices. To get unified \cahk activity index, we re-calculated the $R_{\rm{HK}}^{'}$ of these samples using the methods described in Section \ref{cahk.sec}, but with an integration widow of 1.09 \AA\ for the \cahk lines. We also cross-matched dwarfs from \cite{2023ApJS..264...12H} with the \emph{GALEX} GR6+7 catalog using a radius of 3'' and calculated $R_{\rm{NUV}}^{'}$ of those with NUV detection following Section \ref{nuv.sec}. Meanwhile, we directly gathered $R_{\rm{H\alpha}}^{'}$ and $R_{\rm{NUV}}^{'}$ values of dwarf stars from these studies. 

Finally, we got 181, 125, and 27 giants with $R_{\rm{HK}}^{'}$, $R_{\rm{H\alpha}}^{'}$ and $R_{\rm{NUV}}^{'}$ measurements, respectively.
In addition, we cross-matched our giant sample with the catalog from \cite{2023A&A...675A..26W} to distinguish between red giant branch (RGB) and red clump (RC) stars. Our sample includes 163 RC stars, 40 RGB stars, and 7 giants without classification. The final results together with stellar parameters are presented in Table \ref{tab:table1}. 

Late type stars have convective envelope and radiative core and thus could develop a solar-like dynamo, i.e., $\rm{\alpha}-\rm{\Omega}$ dynamo \citep{1955ApJ...122..293P}. As a result, the well-known activity--rotation relations could act as diagnostics of stellar dynamos. For dwarfs, various chromospheric indices including \cahk lines, $\rm{H\alpha}$ lines and NUV emissions follow a similar relation, including a saturated region and a decay region. \citep[e.g.][]{1984ApJ...279..763N, 2011ApJ...743...48W, 2016ApJ...821...93N}. In this work, we built activity--rotation relations for both dwarfs and evolved stars (Figure \ref{relation}).

In the $R_{\rm{HK}}^{'}-$Ro relation (Figure \ref{relation}) we also included sample from \cite{2020NatAs...4..658L}. Their $S_{\rm{MW}}$ values were converted to $R_{\rm{HK}}^{'}$ using the the method in Section \ref{cahk.sec}. In addition, we removed targets flagged with ``likely spurious $P_{\rm{rot}}$'' and ``unreliable $\tau_{\rm{c}}$'' by \cite{2020NatAs...4..658L}. 
We also classified their sample into single stars and binaries using following steps.
We cross-matched their sample with Gaia eDR3 and the NSS catalog, and selected those with RUWE values larger than 1.4 or listed in the NSS catalog. Furthermore, by cross-matching with SIMBAD, we identified sources marked as spectroscopic binaries or RS Canum Venaticorum variables. 
Most of the remaining single stars from \cite{2020NatAs...4..658L} are subgiants, which follows a similar $R_{\rm{HK}}^{'}-$Ro relation as dwarfs. $R_{\rm{HK}}^{'}$ values of giants exhibit some scatter, which is expected given that either stellar activity or the $B-V$ color can vary, with the latter could influence the calculation of the photospheric \cahk flux.
We remind that the $S-$index of evolved stars in both our sample and \cite{2020NatAs...4..658L} was calculated with a 2 \AA\ window, while the photospheric contribution to the \cahk lines was estimated using a 1 \AA \, window following \cite{2013A&A...549A.117M}. This may overestimate the $R_{\rm{HK}}^{'}$ values of giant stars. 

All the RGB stars, RC stars, and dwarfs fall on a single sequence in the unsaturated regime, and we did not observe a separation between the RGB stars and RC stars in any of the three activity--rotation relations.
We also found that, giants exhibit NUV activity levels similar to those of G and K dwarfs. Although some giants show higher levels compared to M dwarfs, the difference is not as significant as reported by \cite{2020AJ....160...12D}, who reported that the relationship between NUV excess and the Rossby number for evolved stars follows a similar trend to that seen in M dwarfs, but with different slope.
We propose that in evolved stars, the magnetic field is also generated by both convection and rotation, suggesting a shared dynamo mechanism between giants and dwarfs.

Figure \ref{relation_binary} shows a comparison between single and binary evolved stars. We did not find clear difference in magnetic activity levels between single stars and binaries, suggesting that the magnetic activity of these binaries is dominated by the giant component. 
This is not unexpected, as most of these giants in binaries have long periods (i.e., rotational or orbital).
Therefore, we do not observe the results like that giants in close binaries exhibit significantly higher activity levels than single giants, since their rotation rates may have been greatly increased due to tidal interactions with their companions \citep{2022A&A...668A.116G}.

\begin{table*}
\begin{center}
{\footnotesize
  \setlength{\tabcolsep}{1pt}
    \caption{Stellar parameters of and activity proxies of evolved stars in our sample.}
    \label{tab:table1}
    \begin{tabular}{cccccccccccl}
      \toprule 
       Gaia ID & 
       R.A. &
       Decl. &
       $T_{\rm eff}$ &
       $\rm{log\emph{g}}$ &
       [Fe/H] &
       log$R_{\rm{HK}}^{'}$ & 
       log$R_{\rm{H\alpha}}^{'}$ & 
       log$R_{\rm{NUV}}^{'}$ &
       $P_{\rm{rot}}$ &
       $\tau_{c}$ & 
       flag
       \\
       & (degree) & (degree) & (K) & (dex) & (dex) &  & & & (day) & (day)  \\
       (1) & (2) & (3) & (4) & (5) & (6) & (7) & (8) & (9) & (10) & (11)\\
      \hline \\ 

2050241224731793408 & 291.35999 & 36.81989 & 4835$\pm$27 & 2.47$\pm$0.04 & $-$0.45$\pm$0.02 & $-$4.55$\pm$0.09 & $-$4.48$\pm$0.08 & - & 46.74 & 307.54 & RC/b\\
2051086822197774464 & 290.51929 & 37.52545 & 4852$\pm$35 & 2.4$\pm$0.05 & $-$0.5$\pm$0.03 & $-$4.08$\pm$0.07 & $-$4.42$\pm$0.06 & - & 40.98 & 321.54 & RC/s\\
2051739416708443392 & 292.16775 & 37.39372 & 4775$\pm$41 & 2.5$\pm$0.06 & $-$0.34$\pm$0.04 & $-$4.31$\pm$0.19 & $-$4.37$\pm$0.63 & - & 44.54 & 293.96 & RC/s\\
2076974033203960448 & 297.30347 & 42.02882 & 5122$\pm$36 & 3.05$\pm$0.05 & 0.1$\pm$0.03 & $-$4.24$\pm$0.06 & - & $-$2.93$\pm$0.6 & 103.74 & 225.71 & RGB/s\\
2087245121068056064 & 297.54099 & 49.86628 & 5035$\pm$28 & 2.96$\pm$0.04 & 0.17$\pm$0.02 & $-$4.33$\pm$0.07 & $-$4.6$\pm$0.09 & $-$4.43$\pm$1.18 & 97.47 & 219.64 & RC/s\\
2101726238962567552 & 291.22244 & 41.82693 & 4713$\pm$26 & 2.8$\pm$0.04 & $-$0.14$\pm$0.02 & $-$4.23$\pm$0.1 & $-$4.83$\pm$0.65 & $-$3.95$\pm$1.16 & 41.27 & 349.67 & RGB/s\\
2102112339341027840 & 288.37677 & 40.94816 & 4939$\pm$40 & 3.56$\pm$0.07 & 0.22$\pm$0.04 & $-$5.61$\pm$8.26 & - & $-$3.01$\pm$0.08 & 62.11 & 311.86 & RGB/s\\
2102592478023537536 & 287.93837 & 42.92124 & 4901$\pm$20 & 3.03$\pm$0.03 & 0.29$\pm$0.01 & $-$4.2$\pm$0.07 & - & $-$3.66$\pm$0.12 & 120.88 & 398.25 & RC/s\\
2105404311512378496 & 282.38805 & 44.41601 & 4774$\pm$22 & 2.44$\pm$0.03 & $-$0.42$\pm$0.01 & $-$4.63$\pm$0.09 & $-$4.57$\pm$0.07 & - & 101.64 & 294.93 & RC/b\\
2125777987100768512 & 291.77075 & 42.56832 & 4948$\pm$13 & 2.79$\pm$0.02 & $-$0.01$\pm$0.01 & $-$4.01$\pm$0.07 & $-$4.67$\pm$0.12 & $-$3.47$\pm$0.4 & 134.98 & 402.13 & -/s\\

... & ... & ... & ... & ... & ... & ... & ... & ... & ... & ... & ...\\
      \hline \\
    \end{tabular}\\}
  \end{center}
NOTE. (1) Gaia ID: Gaia eDR3 source ID. (2) R.A.: right ascension (3) Decl.: declination. (4) $T_{\rm{eff}}$: effective temperature. (5) log$g$: surface gravity. (6) [Fe/H]: metallicity. (7) log$R_{\rm{HK}}^{'}$ : \cahk proxy. (8) log$R_{\rm{H\alpha}}^{'}$ : $\rm{H\alpha}$ proxy. (9) log$R_{\rm{NUV}}^{'}$ : NUV proxy. (10) $P_{\rm{rot}}$ : rotation period. (11) $\tau_{c}$ : convective turnover time. (12) flag: evolutionary stage. Single stars are marked by `s', potential binaries are marked by `b'.
\end{table*}

\section{summary}

In this study, based on data from the LAMOST and \emph{GALEX} sky surveys, together with the catalog from \cite{2017A&A...605A.111C}, which estimated rotation periods of evolved stars, we investigated the activity--rotation relations for evolved stars using the $R_{\rm{HK}}^{'}$, $R_{\rm{H\alpha}}^{'}$, and $R_{\rm{NUV}}^{'}$ indices.

We found that, for all these three activity indices, evolved stars and dwarfs obey a similar power-law in the unsaturated region of activity--rotation relations, indicating a shared dynamo mechanism in giant and dwarfs.
Our results generally show that the NUV activity levels of giants are comparable to those of G- and K-type dwarfs and are higher than those of M dwarfs. However, the difference is not as significant as reported by \cite{2020AJ....160...12D}. 
Additionally, we did not find significant difference in the activity levels between RC stars and RGB stars.

Furthermore, there is no clear difference in the activity levels between single and binary evolved stars, indicating that the magnetic activity in binaries is dominated by the giant component. 
This is unsurprising, as most of the binaries in our sample are not close-in systems based on their periods, with tidal interactions too weak to enhance magnetic activity.

It is important to note that our $R_{\rm{HK}}^{'}$ values for giants are based on the $S$-index calculated using a 2.18 \AA\ triangle integration window, following the method of the MWO and accounting for the Wilson-Bappu effect, which suggests that the widths of the emission cores of the \cahk lines increase with stellar luminosity. For dwarfs, however, the $S$-index was still calculated using a 1.09 \AA\ window. No corrections were made for the differences in integration windows, which may introduce some uncertainties.

\acknowledgements
We thank the referee for the comprehensive and useful comments. The Guoshoujing Telescope (the Large Sky Area Multi-Object Fiber Spectroscopic Telescope LAMOST) is a National Major Scientific Project built by the Chinese Academy of Sciences. Funding for the project has been provided by the National Development and Reform Commission. LAMOST is operated and managed by the National Astronomical Observatories, Chinese Academy of Sciences.
Some of the data presented in this paper were obtained from the Mikulski Archive for Space Telescopes (MAST).
This work presents results from the European Space Agency (ESA) space mission {\it Gaia}. {\it Gaia} data are being processed by the {\it Gaia} Data Processing and Analysis Consortium (DPAC). Funding for the DPAC is provided by national institutions, in particular the institutions participating in the {\it Gaia} MultiLateral Agreement (MLA). The {\it Gaia} mission website is https://www.cosmos.esa.int/gaia. The {\it Gaia} archive website is https://archives.esac.esa.int/gaia. We acknowledge use of the VizieR catalog access tool, operated at CDS, Strasbourg, France. This work was supported by National Key Research and Development Program of China (NKRDPC) under grant No. 2019YFA0405000, Science Research Grants from the China Manned Space Project with No. CMS-CSST-2021-A08, Strategic Priority Program of the Chinese Academy of Sciences undergrant No. XDB4100000, and National  Natural Science Foundation of China (NSFC) under grant Nos. 11988101/11933004/11833002/12090042/12273057.

\bibliographystyle{yahapj}
\bibliography{main}

\begin{thebibliography}{}
\providecommand\natexlab[1]{#1}
\providecommand\JournalTitle[1]{#1}

\bibitem[{{Ayres} {et~al.}(1981){Ayres}, {Linsky}, {Vaiana}, {Golub}, \&
  {Rosner}}]{1981ApJ...250..293A}
{Ayres}, T.~R., {Linsky}, J.~L., {Vaiana}, G.~S., {Golub}, L., \& {Rosner}, R.
  1981, \href{http://dx.doi.org/10.1086/159374}{\JournalTitle{\apj}, 250, 293}

\bibitem[{{Bailer-Jones} {et~al.}(2021){Bailer-Jones}, {Rybizki}, {Fouesneau},
  {Demleitner}, \& {Andrae}}]{2021AJ....161..147B}
{Bailer-Jones}, C.~A.~L., {Rybizki}, J., {Fouesneau}, M., {Demleitner}, M., \&
  {Andrae}, R. 2021,
  \href{http://dx.doi.org/10.3847/1538-3881/abd806}{\JournalTitle{\aj}, 161,
  147}

\bibitem[{{Baliunas} {et~al.}(1983){Baliunas}, {Hartmann}, \&
  {Dupree}}]{1983ApJ...271..672B}
{Baliunas}, S.~L., {Hartmann}, L., \& {Dupree}, A.~K. 1983,
  \href{http://dx.doi.org/10.1086/161234}{\JournalTitle{\apj}, 271, 672}

\bibitem[{{Barden}(1985)}]{1985ApJ...295..162B}
{Barden}, S.~C. 1985,
  \href{http://dx.doi.org/10.1086/163361}{\JournalTitle{\apj}, 295, 162}

\bibitem[{{Bianchi} {et~al.}(2017){Bianchi}, {Shiao}, \&
  {Thilker}}]{2017ApJS..230...24B}
{Bianchi}, L., {Shiao}, B., \& {Thilker}, D. 2017,
  \href{http://dx.doi.org/10.3847/1538-4365/aa7053}{\JournalTitle{\apjs}, 230,
  24}

\bibitem[{{Boro Saikia} {et~al.}(2018){Boro Saikia}, {Marvin}, {Jeffers},
  {Reiners}, {Cameron}, {Marsden}, {Petit}, {Warnecke}, \&
  {Yadav}}]{2018A&A...616A.108B}
{Boro Saikia}, S., {Marvin}, C.~J., {Jeffers}, S.~V., {et~al.} 2018,
  \href{http://dx.doi.org/10.1051/0004-6361/201629518}{\JournalTitle{\aap},
  616, A108}

\bibitem[{{Bressan} {et~al.}(2012){Bressan}, {Marigo}, {Girardi}, {Salasnich},
  {Dal Cero}, {Rubele}, \& {Nanni}}]{2012MNRAS.427..127B}
{Bressan}, A., {Marigo}, P., {Girardi}, L., {et~al.} 2012,
  \href{http://dx.doi.org/10.1111/j.1365-2966.2012.21948.x}{\JournalTitle{\mnras},
  427, 127}

\bibitem[{{Cardelli} {et~al.}(1989){Cardelli}, {Clayton}, \&
  {Mathis}}]{1989ApJ...345..245C}
{Cardelli}, J.~A., {Clayton}, G.~C., \& {Mathis}, J.~S. 1989,
  \href{http://dx.doi.org/10.1086/167900}{\JournalTitle{\apj}, 345, 245}

\bibitem[{{Ceillier} {et~al.}(2017){Ceillier}, {Tayar}, {Mathur}, {Salabert},
  {Garc{\'\i}a}, {Stello}, {Pinsonneault}, {van Saders}, {Beck}, \&
  {Bloemen}}]{2017A&A...605A.111C}
{Ceillier}, T., {Tayar}, J., {Mathur}, S., {et~al.} 2017,
  \href{http://dx.doi.org/10.1051/0004-6361/201629884}{\JournalTitle{\aap},
  605, A111}

\bibitem[{{Cui} {et~al.}(2012){Cui}, {Zhao}, {Chu}, {Li}, {Li}, {Zhang}, {Su},
  {Yao}, {Wang}, {Xing}, {Li}, {Zhu}, {Wang}, {Gu}, {Luo}, {Xu}, {Zhang},
  {Liu}, {Zhang}, {Yang}, {Cao}, {Chen}, {Chen}, {Chen}, {Chen}, {Chu}, {Feng},
  {Gong}, {Hou}, {Hu}, {Hu}, {Hu}, {Jia}, {Jiang}, {Jiang}, {Jiang}, {Jin},
  {Li}, {Li}, {Li}, {Liu}, {Liu}, {Lu}, {Mao}, {Men}, {Qi}, {Qi}, {Shi},
  {Tang}, {Tao}, {Wang}, {Wang}, {Wang}, {Wang}, {Wang}, {Wang}, {Wang},
  {Wang}, {Wang}, {Wang}, {Wang}, {Wang}, {Xu}, {Xu}, {Yang}, {Yu}, {Yuan},
  {Yuan}, {Zhai}, {Zhang}, {Zhang}, {Zhang}, {Zhao}, {Zhou}, {Zhou}, {Zhu}, \&
  {Zou}}]{2012RAA....12.1197C}
{Cui}, X.-Q., {Zhao}, Y.-H., {Chu}, Y.-Q., {et~al.} 2012,
  \href{http://dx.doi.org/10.1088/1674-4527/12/9/003}{\JournalTitle{Research in
  Astronomy and Astrophysics}, 12, 1197}

\bibitem[{{Dixon} {et~al.}(2020){Dixon}, {Tayar}, \&
  {Stassun}}]{2020AJ....160...12D}
{Dixon}, D., {Tayar}, J., \& {Stassun}, K.~G. 2020,
  \href{http://dx.doi.org/10.3847/1538-3881/ab9080}{\JournalTitle{\aj}, 160,
  12}

\bibitem[{{Duncan} {et~al.}(1991){Duncan}, {Vaughan}, {Wilson}, {Preston},
  {Frazer}, {Lanning}, {Misch}, {Mueller}, {Soyumer}, {Woodard}, {Baliunas},
  {Noyes}, {Hartmann}, {Porter}, {Zwaan}, {Middelkoop}, {Rutten}, \&
  {Mihalas}}]{1991ApJS...76..383D}
{Duncan}, D.~K., {Vaughan}, A.~H., {Wilson}, O.~C., {et~al.} 1991,
  \href{http://dx.doi.org/10.1086/191572}{\JournalTitle{\apjs}, 76, 383}

\bibitem[{{Fitzpatrick}(1999)}]{1999PASP..111...63F}
{Fitzpatrick}, E.~L. 1999,
  \href{http://dx.doi.org/10.1086/316293}{\JournalTitle{\pasp}, 111, 63}

\bibitem[{{Fontenla} {et~al.}(2015){Fontenla}, {Stancil}, \&
  {Landi}}]{2015ApJ...809..157F}
{Fontenla}, J.~M., {Stancil}, P.~C., \& {Landi}, E. 2015,
  \href{http://dx.doi.org/10.1088/0004-637X/809/2/157}{\JournalTitle{\apj},
  809, 157}

\bibitem[{{Fu} {et~al.}(2020){Fu}, {Cat}, {Zong}, {Frasca}, {Gray}, {Ren},
  {Molenda-{\.Z}akowicz}, {Corbally}, {Catanzaro}, {Shi}, {Luo}, \&
  {Zhang}}]{2020RAA....20..167F}
{Fu}, J.-N., {Cat}, P.~D., {Zong}, W., {et~al.} 2020,
  \href{http://dx.doi.org/10.1088/1674-4527/20/10/167}{\JournalTitle{Research
  in Astronomy and Astrophysics}, 20, 167}

\bibitem[{{Gaia Collaboration} {et~al.}(2021){Gaia Collaboration}, {Brown},
  {Vallenari}, {Prusti}, {de Bruijne}, {Babusiaux}, {Biermann}, {Creevey},
  {Evans}, {Eyer}, {Hutton}, {Jansen}, {Jordi}, {Klioner}, {Lammers},
  {Lindegren}, {Luri}, {Mignard}, {Panem}, {Pourbaix}, {Randich}, {Sartoretti},
  {Soubiran}, {Walton}, {Arenou}, {Bailer-Jones}, {Bastian}, {Cropper},
  {Drimmel}, {Katz}, {Lattanzi}, {van Leeuwen}, {Bakker}, {Cacciari},
  {Casta{\~n}eda}, {De Angeli}, {Ducourant}, {Fabricius}, {Fouesneau},
  {Fr{\'e}mat}, {Guerra}, {Guerrier}, {Guiraud}, {Jean-Antoine Piccolo},
  {Masana}, {Messineo}, {Mowlavi}, {Nicolas}, {Nienartowicz}, {Pailler},
  {Panuzzo}, {Riclet}, {Roux}, {Seabroke}, {Sordo}, {Tanga}, {Th{\'e}venin},
  {Gracia-Abril}, {Portell}, {Teyssier}, {Altmann}, {Andrae}, {Bellas-Velidis},
  {Benson}, {Berthier}, {Blomme}, {Brugaletta}, {Burgess}, {Busso}, {Carry},
  {Cellino}, {Cheek}, {Clementini}, {Damerdji}, {Davidson}, {Delchambre},
  {Dell'Oro}, {Fern{\'a}ndez-Hern{\'a}ndez}, {Galluccio}, {Garc{\'\i}a-Lario},
  {Garcia-Reinaldos}, {Gonz{\'a}lez-N{\'u}{\~n}ez}, {Gosset}, {Haigron},
  {Halbwachs}, {Hambly}, {Harrison}, {Hatzidimitriou}, {Heiter},
  {Hern{\'a}ndez}, {Hestroffer}, {Hodgkin}, {Holl}, {Jan{\ss}en}, {Jevardat de
  Fombelle}, {Jordan}, {Krone-Martins}, {Lanzafame}, {L{\"o}ffler}, {Lorca},
  {Manteiga}, {Marchal}, {Marrese}, {Moitinho}, {Mora}, {Muinonen}, {Osborne},
  {Pancino}, {Pauwels}, {Petit}, {Recio-Blanco}, {Richards}, {Riello},
  {Rimoldini}, {Robin}, {Roegiers}, {Rybizki}, {Sarro}, {Siopis}, {Smith},
  {Sozzetti}, {Ulla}, {Utrilla}, {van Leeuwen}, {van Reeven}, {Abbas}, {Abreu
  Aramburu}, {Accart}, {Aerts}, {Aguado}, {Ajaj}, {Altavilla}, {{\'A}lvarez},
  {{\'A}lvarez Cid-Fuentes}, {Alves}, {Anderson}, {Anglada Varela}, {Antoja},
  {Audard}, {Baines}, {Baker}, {Balaguer-N{\'u}{\~n}ez}, {Balbinot}, {Balog},
  {Barache}, {Barbato}, {Barros}, {Barstow}, {Bartolom{\'e}}, {Bassilana},
  {Bauchet}, {Baudesson-Stella}, {Becciani}, {Bellazzini}, {Bernet}, {Bertone},
  {Bianchi}, {Blanco-Cuaresma}, {Boch}, {Bombrun}, {Bossini}, {Bouquillon},
  {Bragaglia}, {Bramante}, {Breedt}, {Bressan}, {Brouillet}, {Bucciarelli},
  {Burlacu}, {Busonero}, {Butkevich}, {Buzzi}, {Caffau}, {Cancelliere},
  {C{\'a}novas}, {Cantat-Gaudin}, {Carballo}, {Carlucci}, {Carnerero},
  {Carrasco}, {Casamiquela}, {Castellani}, {Castro-Ginard}, {Castro Sampol},
  {Chaoul}, {Charlot}, {Chemin}, {Chiavassa}, {Cioni}, {Comoretto}, {Cooper},
  {Cornez}, {Cowell}, {Crifo}, {Crosta}, {Crowley}, {Dafonte}, {Dapergolas},
  {David}, {David}, {de Laverny}, {De Luise}, {De March}, {De Ridder}, {de
  Souza}, {de Teodoro}, {de Torres}, {del Peloso}, {del Pozo}, {Delbo},
  {Delgado}, {Delgado}, {Delisle}, {Di Matteo}, {Diakite}, {Diener},
  {Distefano}, {Dolding}, {Eappachen}, {Edvardsson}, {Enke}, {Esquej}, {Fabre},
  {Fabrizio}, {Faigler}, {Fedorets}, {Fernique}, {Fienga}, {Figueras},
  {Fouron}, {Fragkoudi}, {Fraile}, {Franke}, {Gai}, {Garabato},
  {Garcia-Gutierrez}, {Garc{\'\i}a-Torres}, {Garofalo}, {Gavras}, {Gerlach},
  {Geyer}, {Giacobbe}, {Gilmore}, {Girona}, {Giuffrida}, {Gomel}, {Gomez},
  {Gonzalez-Santamaria}, {Gonz{\'a}lez-Vidal}, {Granvik},
  {Guti{\'e}rrez-S{\'a}nchez}, {Guy}, {Hauser}, {Haywood}, {Helmi}, {Hidalgo},
  {Hilger}, {H{\l}adczuk}, {Hobbs}, {Holland}, {Huckle}, {Jasniewicz},
  {Jonker}, {Juaristi Campillo}, {Julbe}, {Karbevska}, {Kervella}, {Khanna},
  {Kochoska}, {Kontizas}, {Kordopatis}, {Korn}, {Kostrzewa-Rutkowska},
  {Kruszy{\'n}ska}, {Lambert}, {Lanza}, {Lasne}, {Le Campion}, {Le Fustec},
  {Lebreton}, {Lebzelter}, {Leccia}, {Leclerc}, {Lecoeur-Taibi}, {Liao},
  {Licata}, {Lindstr{\o}m}, {Lister}, {Livanou}, {Lobel}, {Madrero Pardo},
  {Managau}, {Mann}, {Marchant}, {Marconi}, {Marcos Santos}, {Marinoni},
  {Marocco}, {Marshall}, {Martin Polo}, {Mart{\'\i}n-Fleitas}, {Masip},
  {Massari}, {Mastrobuono-Battisti}, {Mazeh}, {McMillan}, {Messina},
  {Michalik}, {Millar}, {Mints}, {Molina}, {Molinaro}, {Moln{\'a}r},
  {Montegriffo}, {Mor}, {Morbidelli}, {Morel}, {Morris}, {Mulone}, {Munoz},
  {Muraveva}, {Murphy}, {Musella}, {Noval}, {Ord{\'e}novic}, {Orr{\`u}},
  {Osinde}, {Pagani}, {Pagano}, {Palaversa}, {Palicio}, {Panahi}, {Pawlak},
  {Pe{\~n}alosa Esteller}, {Penttil{\"a}}, {Piersimoni}, {Pineau}, {Plachy},
  {Plum}, {Poggio}, {Poretti}, {Poujoulet}, {Pr{\v{s}}a}, {Pulone}, {Racero},
  {Ragaini}, {Rainer}, {Raiteri}, {Rambaux}, {Ramos}, {Ramos-Lerate}, {Re
  Fiorentin}, {Regibo}, {Reyl{\'e}}, {Ripepi}, {Riva}, {Rixon}, {Robichon},
  {Robin}, {Roelens}, {Rohrbasser}, {Romero-G{\'o}mez}, {Rowell}, {Royer},
  {Rybicki}, {Sadowski}, {Sagrist{\`a} Sell{\'e}s}, {Sahlmann}, {Salgado},
  {Salguero}, {Samaras}, {Sanchez Gimenez}, {Sanna}, {Santove{\~n}a},
  {Sarasso}, {Schultheis}, {Sciacca}, {Segol}, {Segovia}, {S{\'e}gransan},
  {Semeux}, {Shahaf}, {Siddiqui}, {Siebert}, {Siltala}, {Slezak}, {Smart},
  {Solano}, {Solitro}, {Souami}, {Souchay}, {Spagna}, {Spoto}, {Steele},
  {Steidelm{\"u}ller}, {Stephenson}, {S{\"u}veges}, {Szabados}, {Szegedi-Elek},
  {Taris}, {Tauran}, {Taylor}, {Teixeira}, {Thuillot}, {Tonello}, {Torra},
  {Torra}, {Turon}, {Unger}, {Vaillant}, {van Dillen}, {Vanel}, {Vecchiato},
  {Viala}, {Vicente}, {Voutsinas}, {Weiler}, {Wevers}, {Wyrzykowski}, {Yoldas},
  {Yvard}, {Zhao}, {Zorec}, {Zucker}, {Zurbach}, \&
  {Zwitter}}]{2021A&A...649A...1G}
{Gaia Collaboration}, {Brown}, A.~G.~A., {Vallenari}, A., {et~al.} 2021,
  \href{http://dx.doi.org/10.1051/0004-6361/202039657}{\JournalTitle{\aap},
  649, A1}

\bibitem[{{Gaulme} {et~al.}(2020){Gaulme}, {Jackiewicz}, {Spada}, {Chojnowski},
  {Mosser}, {McKeever}, {Hedlund}, {Vrard}, {Benbakoura}, \&
  {Damiani}}]{2020A&A...639A..63G}
{Gaulme}, P., {Jackiewicz}, J., {Spada}, F., {et~al.} 2020,
  \href{http://dx.doi.org/10.1051/0004-6361/202037781}{\JournalTitle{\aap},
  639, A63}

\bibitem[{{Gehan} {et~al.}(2022){Gehan}, {Gaulme}, \&
  {Yu}}]{2022A&A...668A.116G}
{Gehan}, C., {Gaulme}, P., \& {Yu}, J. 2022,
  \href{http://dx.doi.org/10.1051/0004-6361/202245083}{\JournalTitle{\aap},
  668, A116}

\bibitem[{{Green}(2018)}]{2018JOSS....3..695M}
{Green}, G. 2018,
  \href{http://dx.doi.org/10.21105/joss.00695}{\JournalTitle{The Journal of
  Open Source Software}, 3, 695}

\bibitem[{{Green} {et~al.}(2019){Green}, {Schlafly}, {Zucker}, {Speagle}, \&
  {Finkbeiner}}]{2019ApJ...887...93G}
{Green}, G.~M., {Schlafly}, E., {Zucker}, C., {Speagle}, J.~S., \&
  {Finkbeiner}, D. 2019,
  \href{http://dx.doi.org/10.3847/1538-4357/ab5362}{\JournalTitle{\apj}, 887,
  93}

\bibitem[{{Hall} {et~al.}(2007){Hall}, {Lockwood}, \&
  {Skiff}}]{2007AJ....133..862H}
{Hall}, J.~C., {Lockwood}, G.~W., \& {Skiff}, B.~A. 2007,
  \href{http://dx.doi.org/10.1086/510356}{\JournalTitle{\aj}, 133, 862}

\bibitem[{{Han} {et~al.}(2023){Han}, {Wang}, {Bai}, {Yang}, {Fang}, \&
  {Liu}}]{2023ApJS..264...12H}
{Han}, H., {Wang}, S., {Bai}, Y., {et~al.} 2023,
  \href{http://dx.doi.org/10.3847/1538-4365/ac9eac}{\JournalTitle{\apjs}, 264,
  12}

\bibitem[{{Han} {et~al.}(2024){Han}, {Wang}, {Zheng}, {Li}, {Xiao}, \&
  {Liu}}]{2024ApJS..273....8H}
{Han}, H., {Wang}, S., {Zheng}, C., {et~al.} 2024,
  \href{http://dx.doi.org/10.3847/1538-4365/ad4b17}{\JournalTitle{\apjs}, 273,
  8}

\bibitem[{{Husser} {et~al.}(2013){Husser}, {Wende-von Berg}, {Dreizler},
  {Homeier}, {Reiners}, {Barman}, \& {Hauschildt}}]{2013A&A...553A...6H}
{Husser}, T.~O., {Wende-von Berg}, S., {Dreizler}, S., {et~al.} 2013,
  \href{http://dx.doi.org/10.1051/0004-6361/201219058}{\JournalTitle{\aap},
  553, A6}

\bibitem[{{Karoff} {et~al.}(2016){Karoff}, {Knudsen}, {De Cat}, {Bonanno},
  {Fogtmann-Schulz}, {Fu}, {Frasca}, {Inceoglu}, {Olsen}, {Zhang}, {Hou},
  {Wang}, {Shi}, \& {Zhang}}]{2016NatCo...711058K}
{Karoff}, C., {Knudsen}, M.~F., {De Cat}, P., {et~al.} 2016,
  \href{http://dx.doi.org/10.1038/ncomms11058}{\JournalTitle{Nature
  Communications}, 7, 11058}

\bibitem[{{Koch} {et~al.}(2010){Koch}, {Borucki}, {Basri}, {Batalha}, {Brown},
  {Caldwell}, {Christensen-Dalsgaard}, {Cochran}, {DeVore}, {Dunham},
  {Gautier}, {Geary}, {Gilliland}, {Gould}, {Jenkins}, {Kondo}, {Latham},
  {Lissauer}, {Marcy}, {Monet}, {Sasselov}, {Boss}, {Brownlee}, {Caldwell},
  {Dupree}, {Howell}, {Kjeldsen}, {Meibom}, {Morrison}, {Owen}, {Reitsema},
  {Tarter}, {Bryson}, {Dotson}, {Gazis}, {Haas}, {Kolodziejczak}, {Rowe}, {Van
  Cleve}, {Allen}, {Chandrasekaran}, {Clarke}, {Li}, {Quintana}, {Tenenbaum},
  {Twicken}, \& {Wu}}]{2010ApJ...713L..79K}
{Koch}, D.~G., {Borucki}, W.~J., {Basri}, G., {et~al.} 2010,
  \href{http://dx.doi.org/10.1088/2041-8205/713/2/L79}{\JournalTitle{\apjl},
  713, L79}

\bibitem[{{Lehtinen} {et~al.}(2020){Lehtinen}, {Spada}, {K{\"a}pyl{\"a}},
  {Olspert}, \& {K{\"a}pyl{\"a}}}]{2020NatAs...4..658L}
{Lehtinen}, J.~J., {Spada}, F., {K{\"a}pyl{\"a}}, M.~J., {Olspert}, N., \&
  {K{\"a}pyl{\"a}}, P.~J. 2020,
  \href{http://dx.doi.org/10.1038/s41550-020-1039-x}{\JournalTitle{Nature
  Astronomy}, 4, 658}

\bibitem[{{Li} {et~al.}(2024){Li}, {Wang}, {Han}, {Yang}, {Zheng}, {Huang}, \&
  {Liu}}]{2024ApJ...966...69L}
{Li}, X., {Wang}, S., {Han}, H., {et~al.} 2024,
  \href{http://dx.doi.org/10.3847/1538-4357/ad3038}{\JournalTitle{\apj}, 966,
  69}

\bibitem[{{Linsky}(2017)}]{2017ARA&A..55..159L}
{Linsky}, J.~L. 2017,
  \href{http://dx.doi.org/10.1146/annurev-astro-091916-055327}{\JournalTitle{\araa},
  55, 159}

\bibitem[{{Linsky} \& {Haisch}(1979)}]{1979ApJ...229L..27L}
{Linsky}, J.~L., \& {Haisch}, B.~M. 1979,
  \href{http://dx.doi.org/10.1086/182924}{\JournalTitle{\apjl}, 229, L27}

\bibitem[{{Linsky} {et~al.}(1979){Linsky}, {Worden}, {McClintock}, \&
  {Robertson}}]{1979ApJS...41...47L}
{Linsky}, J.~L., {Worden}, S.~P., {McClintock}, W., \& {Robertson}, R.~M. 1979,
  \href{http://dx.doi.org/10.1086/190607}{\JournalTitle{\apjs}, 41, 47}

\bibitem[{{Luo} {et~al.}(2015){Luo}, {Zhao}, {Zhao}, {Deng}, {Liu}, {Jing},
  {Wang}, {Zhang}, {Shi}, {Cui}, {Chu}, {Li}, {Bai}, {Wu}, {Cai}, {Cao}, {Cao},
  {Carlin}, {Chen}, {Chen}, {Chen}, {Chen}, {Chen}, {Chen}, {Chen},
  {Christlieb}, {Chu}, {Cui}, {Dong}, {Du}, {Fan}, {Feng}, {Fu}, {Gao}, {Gong},
  {Gu}, {Guo}, {Han}, {He}, {Hou}, {Hou}, {Hou}, {Hu}, {Hu}, {Hu}, {Huo},
  {Jia}, {Jiang}, {Jiang}, {Jiang}, {Jin}, {Kong}, {Kong}, {Lei}, {Li}, {Li},
  {Li}, {Li}, {Li}, {Li}, {Li}, {Li}, {Li}, {Li}, {Li}, {Li}, {Liang}, {Lin},
  {Liu}, {Liu}, {Liu}, {Liu}, {Lu}, {Luo}, {Mao}, {Newberg}, {Ni}, {Qi}, {Qi},
  {Shen}, {Shi}, {Song}, {Song}, {Su}, {Su}, {Tang}, {Tao}, {Tian}, {Wang},
  {Wang}, {Wang}, {Wang}, {Wang}, {Wang}, {Wang}, {Wang}, {Wang}, {Wang},
  {Wang}, {Wang}, {Wang}, {Wang}, {Wang}, {Wang}, {Wang}, {Wang}, {Wang},
  {Wang}, {Wei}, {Wei}, {Wu}, {Wu}, {Wu}, {Wu}, {Xing}, {Xu}, {Xu}, {Xu},
  {Yan}, {Yang}, {Yang}, {Yang}, {Yang}, {Yao}, {Yu}, {Yuan}, {Yuan}, {Yuan},
  {Yuan}, {Zhai}, {Zhang}, {Zhang}, {Zhang}, {Zhang}, {Zhang}, {Zhang},
  {Zhang}, {Zhang}, {Zhao}, {Zhou}, {Zhou}, {Zhu}, {Zhu}, {Zou}, \&
  {Zuo}}]{2015RAA....15.1095L}
{Luo}, A.~L., {Zhao}, Y.-H., {Zhao}, G., {et~al.} 2015,
  \href{http://dx.doi.org/10.1088/1674-4527/15/8/002}{\JournalTitle{Research in
  Astronomy and Astrophysics}, 15, 1095}

\bibitem[{{Middelkoop}(1982)}]{1982A&A...113....1M}
{Middelkoop}, F. 1982, \JournalTitle{\aap}, 113, 1

\bibitem[{{Mittag} {et~al.}(2013){Mittag}, {Schmitt}, \&
  {Schr{\"o}der}}]{2013A&A...549A.117M}
{Mittag}, M., {Schmitt}, J.~H.~M.~M., \& {Schr{\"o}der}, K.~P. 2013,
  \href{http://dx.doi.org/10.1051/0004-6361/201219868}{\JournalTitle{\aap},
  549, A117}

\bibitem[{{Morrissey} {et~al.}(2007){Morrissey}, {Conrow}, {Barlow}, {Small},
  {Seibert}, {Wyder}, {Budav{\'a}ri}, {Arnouts}, {Friedman}, {Forster},
  {Martin}, {Neff}, {Schiminovich}, {Bianchi}, {Donas}, {Heckman}, {Lee},
  {Madore}, {Milliard}, {Rich}, {Szalay}, {Welsh}, \&
  {Yi}}]{2007ApJS..173..682M}
{Morrissey}, P., {Conrow}, T., {Barlow}, T.~A., {et~al.} 2007,
  \href{http://dx.doi.org/10.1086/520512}{\JournalTitle{\apjs}, 173, 682}

\bibitem[{{Morton}(2015)}]{2015ascl.soft03010M}
{Morton}, T.~D. 2015, {isochrones: Stellar model grid package}, Astrophysics
  Source Code Library, record ascl:1503.010

\bibitem[{{Newton} {et~al.}(2016){Newton}, {Irwin}, {Charbonneau},
  {Berta-Thompson}, {Dittmann}, \& {West}}]{2016ApJ...821...93N}
{Newton}, E.~R., {Irwin}, J., {Charbonneau}, D., {et~al.} 2016,
  \href{http://dx.doi.org/10.3847/0004-637X/821/2/93}{\JournalTitle{\apj}, 821,
  93}

\bibitem[{{Noyes} {et~al.}(1984){Noyes}, {Hartmann}, {Baliunas}, {Duncan}, \&
  {Vaughan}}]{1984ApJ...279..763N}
{Noyes}, R.~W., {Hartmann}, L.~W., {Baliunas}, S.~L., {Duncan}, D.~K., \&
  {Vaughan}, A.~H. 1984,
  \href{http://dx.doi.org/10.1086/161945}{\JournalTitle{\apj}, 279, 763}

\bibitem[{{Parker}(1955)}]{1955ApJ...122..293P}
{Parker}, E.~N. 1955,
  \href{http://dx.doi.org/10.1086/146087}{\JournalTitle{\apj}, 122, 293}

\bibitem[{{Schr{\"o}der} {et~al.}(2018){Schr{\"o}der}, {Schmitt}, {Mittag},
  {G{\'o}mez Trejo}, \& {Jack}}]{2018MNRAS.480.2137S}
{Schr{\"o}der}, K.~P., {Schmitt}, J.~H.~M.~M., {Mittag}, M., {G{\'o}mez Trejo},
  V., \& {Jack}, D. 2018,
  \href{http://dx.doi.org/10.1093/mnras/sty1942}{\JournalTitle{\mnras}, 480,
  2137}

\bibitem[{{Simon} \& {Drake}(1989)}]{1989ApJ...346..303S}
{Simon}, T., \& {Drake}, S.~A. 1989,
  \href{http://dx.doi.org/10.1086/168012}{\JournalTitle{\apj}, 346, 303}

\bibitem[{{Skumanich}(1972)}]{1972ApJ...171..565S}
{Skumanich}, A. 1972,
  \href{http://dx.doi.org/10.1086/151310}{\JournalTitle{\apj}, 171, 565}

\bibitem[{{Spada} {et~al.}(2017){Spada}, {Demarque}, {Kim}, {Boyajian}, \&
  {Brewer}}]{2017ApJ...838..161S}
{Spada}, F., {Demarque}, P., {Kim}, Y.~C., {Boyajian}, T.~S., \& {Brewer},
  J.~M. 2017,
  \href{http://dx.doi.org/10.3847/1538-4357/aa661d}{\JournalTitle{\apj}, 838,
  161}

\bibitem[{{Strassmeier} {et~al.}(1994){Strassmeier}, {Handler}, {Paunzen}, \&
  {Rauth}}]{1994A&A...281..855S}
{Strassmeier}, K.~G., {Handler}, G., {Paunzen}, E., \& {Rauth}, M. 1994,
  \JournalTitle{\aap}, 281, 855

\bibitem[{{STScI}(2013)}]{https://doi.org/10.17909/t9h59d}
{STScI}. 2013, GALEX/MCAT

\bibitem[{{Tayar} {et~al.}(2015){Tayar}, {Ceillier},
  {Garc{\'\i}a-Hern{\'a}ndez}, {Troup}, {Mathur}, {Garc{\'\i}a}, {Zamora},
  {Johnson}, {Pinsonneault}, {M{\'e}sz{\'a}ros}, {Allende Prieto}, {Chaplin},
  {Elsworth}, {Hekker}, {Nidever}, {Salabert}, {Schneider}, {Serenelli},
  {Shetrone}, \& {Stello}}]{2015ApJ...807...82T}
{Tayar}, J., {Ceillier}, T., {Garc{\'\i}a-Hern{\'a}ndez}, D.~A., {et~al.} 2015,
  \href{http://dx.doi.org/10.1088/0004-637X/807/1/82}{\JournalTitle{\apj}, 807,
  82}

\bibitem[{{Van Cleve} \& {Caldwell}(2016)}]{2016ksci.rept....1V}
{Van Cleve}, J.~E., \& {Caldwell}, D.~A. 2016, {Kepler Instrument Handbook},
  Kepler Science Document KSCI-19033-002, id.1. Edited by Michael R. Haas and
  Steve B. Howell

\bibitem[{{Vernazza} {et~al.}(1981){Vernazza}, {Avrett}, \&
  {Loeser}}]{1981ApJS...45..635V}
{Vernazza}, J.~E., {Avrett}, E.~H., \& {Loeser}, R. 1981,
  \href{http://dx.doi.org/10.1086/190731}{\JournalTitle{\apjs}, 45, 635}

\bibitem[{{Walkowicz} {et~al.}(2004){Walkowicz}, {Hawley}, \&
  {West}}]{2004PASP..116.1105W}
{Walkowicz}, L.~M., {Hawley}, S.~L., \& {West}, A.~A. 2004,
  \href{http://dx.doi.org/10.1086/426792}{\JournalTitle{\pasp}, 116, 1105}

\bibitem[{{Wang} {et~al.}(2023){Wang}, {Huang}, {Zhou}, \&
  {Zhang}}]{2023A&A...675A..26W}
{Wang}, C., {Huang}, Y., {Zhou}, Y., \& {Zhang}, H. 2023,
  \href{http://dx.doi.org/10.1051/0004-6361/202245809}{\JournalTitle{\aap},
  675, A26}

\bibitem[{{Wang} {et~al.}(2020){Wang}, {Bai}, {He}, \&
  {Liu}}]{2020ApJ...902..114W}
{Wang}, S., {Bai}, Y., {He}, L., \& {Liu}, J. 2020,
  \href{http://dx.doi.org/10.3847/1538-4357/abb66d}{\JournalTitle{\apj}, 902,
  114}

\bibitem[{{Wilson}(1978)}]{1978ApJ...226..379W}
{Wilson}, O.~C. 1978,
  \href{http://dx.doi.org/10.1086/156618}{\JournalTitle{\apj}, 226, 379}

\bibitem[{{Wilson} \& {Vainu Bappu}(1957)}]{1957ApJ...125..661W}
{Wilson}, O.~C., \& {Vainu Bappu}, M.~K. 1957,
  \href{http://dx.doi.org/10.1086/146339}{\JournalTitle{\apj}, 125, 661}

\bibitem[{{Wright} {et~al.}(2011){Wright}, {Drake}, {Mamajek}, \&
  {Henry}}]{2011ApJ...743...48W}
{Wright}, N.~J., {Drake}, J.~J., {Mamajek}, E.~E., \& {Henry}, G.~W. 2011,
  \href{http://dx.doi.org/10.1088/0004-637X/743/1/48}{\JournalTitle{\apj}, 743,
  48}

\bibitem[{{Zacharias} {et~al.}(2013){Zacharias}, {Finch}, {Girard}, {Henden},
  {Bartlett}, {Monet}, \& {Zacharias}}]{2013AJ....145...44Z}
{Zacharias}, N., {Finch}, C.~T., {Girard}, T.~M., {et~al.} 2013,
  \href{http://dx.doi.org/10.1088/0004-6256/145/2/44}{\JournalTitle{\aj}, 145,
  44}

\bibitem[{{Zhang} {et~al.}(2020{\natexlab{a}}){Zhang}, {Liu}, \&
  {Deng}}]{2020ApJS..246....9Z}
{Zhang}, B., {Liu}, C., \& {Deng}, L.-C. 2020{\natexlab{a}},
  \href{http://dx.doi.org/10.3847/1538-4365/ab55ef}{\JournalTitle{\apjs}, 246,
  9}

\bibitem[{{Zhang} {et~al.}(2020{\natexlab{b}}){Zhang}, {Bi}, {Li}, {Jiang},
  {Li}, {He}, {Yu}, {Khanna}, {Ge}, {Liu}, {Tian}, {Wu}, \&
  {Zhang}}]{2020ApJS..247....9Z}
{Zhang}, J., {Bi}, S., {Li}, Y., {et~al.} 2020{\natexlab{b}},
  \href{http://dx.doi.org/10.3847/1538-4365/ab6165}{\JournalTitle{\apjs}, 247,
  9}

\end{thebibliography}
\clearpage

\end{CJK*}
\end{document}